\documentclass[]{sig-alternate-10pt}
\usepackage[T1]{fontenc}
\usepackage{ucs}
\usepackage[utf8x]{inputenc}
\usepackage{amssymb,amsmath} 
\usepackage{float,cite,comment}
\usepackage{booktabs}
\usepackage{url}
\usepackage{algorithm}
\usepackage{enumitem}
\usepackage{xcolor}
\usepackage{ifpdf}

\usepackage{graphicx}
\ifpdf
    \usepackage{hyperref}
    \hypersetup{pdftitle={\@title},pdfauthor={\@author},pdfsubject={},pdfkeywords={}}
    \DeclareGraphicsExtensions{.jpg,.pdf,.mps,.png}
\else
    \usepackage{hyperref}
    \DeclareGraphicsExtensions{.eps,.ps,.bmp,.tif,.tiff,.tga}
\fi

\newcommand{\censure}[1]{}
\newcommand{\fakesubsection}[1]{~\\\noindent{\textbf{#1}}} 

\newcommand{\C}{$\checkmark$}

\title{A First Look at Anycast CDN Traffic}

\numberofauthors{1}
\author{\alignauthor Danilo Cicalese$^\star$, Danilo Giordano$^\dagger$, Alessandro Finamore$^\ddagger$, Marco Mellia$^\dagger$,\\Maurizio Munaf\`o$^\dagger$, Dario Rossi$^\star$, Diana Joumblatt$^\star$\\\vspace{0.2cm}
\affaddr{$^\star$Telecom Paristech} \hspace{1cm} \affaddr{$^\dagger$Politecnico di Torino} \hspace{1cm}
\affaddr{$^\ddagger$Telef\'onica Research} \\\vspace{0.2cm}
}

\begin{document}
\maketitle

\begin{abstract}
IP anycast routes packets to the topologically nearest server according to BGP proximity. UDP-based services (e.g., DNS resolvers and multicast rendez-vous points),  which are based on a single request-response scheme, have been historically the first to use IP anycast. While there is a common belief in the Internet measurements community that stateful services cannot run on top of anycast due to Internet path instabilities, in this work we shed some light on the usage of anycast by Anycast-enabled Content Delivery Networks (A-CDNs). Indeed, to the best of our knowledge, little is known in the literature about the nature and the dynamics of these new players.%

In this paper, we provide a first look at the traffic of A-CDNs. Our methodology combines active and passive measurements. Building upon our previous work, we use active measurements to detect anycast usage for servers hosting popular websites, and geolocate the A-CDN caches. Next, we characterise the traffic towards A-CDNs in the wild using a month-long dataset collected passively from a European ISP. We find that i)  A-CDNs
such as CloudFlare and EdgeCast serve popular web content over TCP, ii) A-CDN servers are contacted by users on a daily basis, and ii) routes to A-CDN servers are stable with few changes observed. 
\end{abstract}

\section{Introduction}
IP anycast allows a group of geographically distributed servers to share a common IP address. BGP proximity routes traffic to the topologically nearest server. Historically, anycast usage has been restricted to stateless UDP services such as DNS root and top level domain servers,  6-to-4 relay routers, multicast rendezvous points, and sinkholes. Recently, we are witnessing the use of anycast with stateful (TCP) Internet services. In particular, Anycast-enabled Content Delivery Networks (A-CDNs) with geographically large footprints are serving web content using anycast IP addresses for servers. While traditional CDNs rely on DNS- or HTTP-based redirection mechanisms to direct client requests to the nearest cache~\cite{Nygren:akamai,Finamore:youtube}, A-CDNs rely on IP anycast to select the nearest cache and to perform load-balancing among the different caches.

Examples of anycast adoption includes both generic CDN providers like CloudFlare or EdgeCast (recently bought by Verizon), and dedicated deployments such as the Microsoft A-CDN, which serves \emph{bing.com} and \emph{live.com} content.
While some A-CDNs openly disclose the location of their caches~\cite{CF_map,EC_map} and their status~\cite{CF_status,EC_status}, little is known about the volume of traffic they attract, the services they host, and the performance stability they guarantee.  

In this paper, we tackle the \emph{detection} of A-CDNs, and the \emph{characterisation} of the traffic towards them. We use complementary methodologies, leveraging active measurements for detection, and passive measurements for characterisation. Since our goal is to get a conservative yet representative view of anycast usage in the Internet, we map the Alexa top-100k most popular websites to IP/24 subnets, and identify 328  /24 A-CDN subnets that use IP anycast. We exploit the recently developed anycast detection technique~\cite{anycast_geoloc} to geolocate the servers whose addresses belong to anycast subnets. Then, to provide a first characterisation of modern usage of A-CDNs, we use traffic traces from 20,000 households collected from a large European ISP for the entire month of September 2014. In particular, we quantify the volume of traffic towards those A-CDNs, and study the path stability between clients and A-CDN caches.  


We summarise our main findings as follows:
\begin{enumerate}
\item A-CDNs today are a reality and host popular services. In our dataset, we observe 3\% of web traffic towards A-CDNs. In addition, approximately 50\% of users encounter an A-CDN cache during normal browsing activity.
\item Given the relatively small volume of traffic A-CDNs have to handle, A-CDNs have a small geographical footprint in comparison with traditional CDNs. 
\item Internet paths between A-CDNs and clients are stable. The EdgeCast A-CDN did not witness any routing changes during the entire month, while traffic for other A-CDNs revealed few routing events, separated by days of stable configurations. Compared to the typical hourly changes observed in traditional CDNs~\cite{Finamore:youtube,finamore_publicdns}, the association between clients and anycast caches is relatively stable.
\end{enumerate}

In the remainder of this paper, we first discuss our contributions with respect to the literature on anycast and CDNs (Sec.~\ref{sec:related}). Then, we present the results of our active measurements and quantify anycast adoption by the CDNs supporting the top-100k websites from Alexa (Sec.~\ref{sec:active}). In addition, we investigate the properties of A-CDNs traffic (Sec.~\ref{sec:passive}) and the stability of routes towards A-CDN servers (Sec.~\ref{sec:affinity}). Finally, we conclude with a discussion of open issues (Sec.~\ref{sec:open}).

\section{Related work}\label{sec:related}
A large body of work in the literature investigates the impact of anycast usage on service performance by measuring server proximity \cite{LiuHFBC07,SaratPT05, ballani:2005, BallaniFR06,k-root2}, client-server affinity \cite{dns_anycast_stability, LiuHFBC07, SaratPT05, ballani:2005, BallaniFR06,j-root2}, server availability \cite{SaratPT05, BallaniFR06, anycast_bgp}, and load-balancing \cite{BallaniFR06, j-root2}. Several studies~\cite{ballani:2005, Freedman:2006,GIA} propose architectural improvements to address the performance shortcomings of IP anycast in terms of scalability and server selection. More recently, there has been a renewed interest in IP anycast and particularly in techniques to detect anycast usage~\cite{anycasters},  and to enumerate~\cite{FanHG13} and geolocate~\cite{anycast_geoloc} anycast replicas. While in~\cite{anycast_geoloc} the focus is only on DNS servers, in this work, we apply the same anycast enumeration and geolocation technique to form an initial census of anycast IP addresses serving web traffic.


Closest to our work are the studies that investigate client-server affinity and quantify how often packets from a given client reach the same anycast server. Previous efforts studied affinity either by periodically sending probes to anycast addresses and counting server switches \cite{SaratPT05, BallaniFR06, ballani:2005,dns_anycast_stability,anycast_bgp}, or by inspecting traffic at the anycast servers themselves and counting, for each client IP address, the number of times this IP shows up in multiple servers \cite{j-root2, LiuHFBC07}. With the exception of two studies \cite{j-root,dns_anycast_stability}, previous efforts showed that anycast witnesses rare server switching and maintains good connection affinity \cite{k-root, ballani:2005, LiuHFBC07, BallaniFR06, k-root2}. Consequently, stateful services could run on top of anycast. Yet, most of the existing studies \cite{ LiuHFBC07,SaratPT05, ballani:2005,  BallaniFR06,k-root2,dns_anycast_stability, j-root2, anycast_bgp}  evaluate the performance of anycast with UDP services such as DNS root and \texttt{.org} top level domain~\cite{rfc3258,dns_ipanycast,root-servers,Abley:2004:f-root,as112}. One exception is the work of Levine et al. \cite{TCPanycast} which reports positive results from operational experience of running TCP with anycast in CacheFly \cite{cachefly}. CacheFly is the first CDN company to use TCP Anycast. In this paper, we reappraise these results with other popular A-CDNs we find in the wild. To the best of our knowledge, no work in the literature has documented and studied the adoption of anycast by CDN providers. We are thus the first to provide a first look at A-CDN in the Internet.


%
%
\begin{figure}[t]
	\includegraphics[width=\columnwidth]{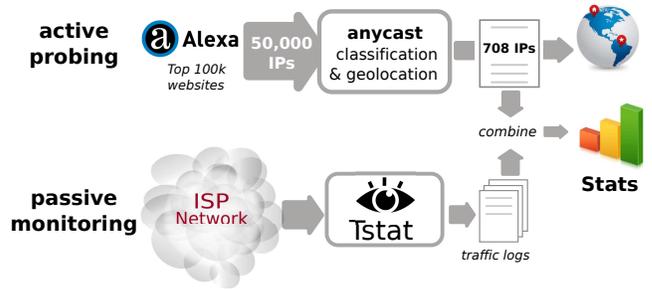}
	\caption{Analysis workflow.\label{fig:workflow}}
\end{figure}

\section{Active Detection Methodology}\label{sec:active}

In this section, we describe the workflow of the active measurement methodology used to detect A-CDNs as schetched in Fig.~\ref{fig:workflow}. First, we compile a list of the top-100k Alexa websites. From each URL, we extract the hostname, and resolve it to IP/32 addresses. We obtain a list of 97,530 unique IP/32 addresses that belong to 50,882 IP/24 subnets. We simultaneously ping all the IP/32 addresses from 250 PlanetLab nodes (a single ICMP sample per-VP, per-IP/32 for a total of 12.7M pings). Next, we ran the anycast detection technique developed in~\cite{anycast_geoloc} to identify IP/32 anycast addresses (i.e., located in more than one geographical location). Due to lack of space, we defer the reader to~\cite{anycast_geoloc} for more details about the anycast detection technique.
In a nutshell, over these collected measurements, we iteratively run a greedy solver for an optimisation problem to verify if the given IP/32 violates the speed of light constraint: when pinged from two different places, the sum of the RTT cannot be smaller than the time light has to spend to go from one probe to the other, i.e., the physics of the triangular inequality must hold. Based then on triangulation, the IP/32 address then is geolocated.
The whole process requires less than 3 minutes to complete. This maximises the probability of completing a census during a stationary period of time.

\begin{figure}[t]
    \begin{center}
        \includegraphics[width=0.45\textwidth]{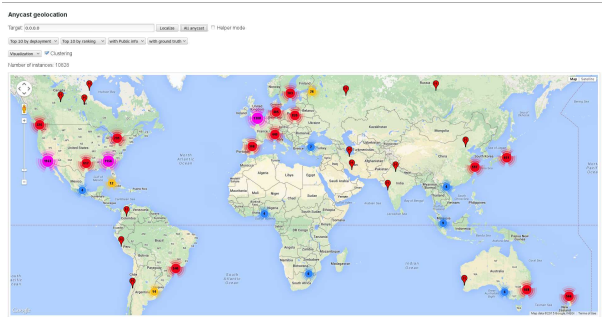}
\caption{Map of IP/24 A-CDN caches. Online data available at~\cite{infocom_demo_website}}\label{fig:demo} 
    \end{center}
\end{figure}

We get a list of 708 IP/32 addresses (328 distinct IP/24 subnets) that are anycast and geolocated within a 300km radius area. These addresses belong to 64 ASes. Notably, three among the top-100 Alexa worldwide ranking are present: \emph{thepiratebay.se}, \emph{reddit.com} both hosted by CloudFlare~\cite{CF_map} and \emph{wordpress.com}  hosted by AUTOMATTIC~\cite{automattic}. The website in~\cite{infocom_demo_website} provides a web-interface that allows the research community to explore our results and in particular the geographical locations of replicas for the anycast IPs identified in the top-100k Alexa ranking. Fig.~\ref{fig:demo} shows a snapshot of the website. More precisely, it provides an aggregate view of the geographical footprints of discovered A-CDNs. Interestingly, this dataset is valuable since it reveals IP/24 anycast addresses belonging to more than 67 organisations, including EdgeCast, CloudFlare, Google and Microsoft.

%
%

\subsection{Discussion}
We discuss some issues related to our measurement choices and results here. 
\noindent
\fakesubsection{Consistency:} Given that IP anycast is based on BGP, it is reasonable to assume that all IP/32 addresses belonging on an anycast IP/24 are also anycast. Previous work~\cite{anycasters} shows that 88\% of the anycast prefixes are /24 or bigger. To confirm this assumption, we run measurements for all IP/32 addresses of a subset of the anycast /24 subnets and obtain results in agreement with this assumption.
\noindent
\fakesubsection{Representativeness.} By restraining ourselves to the top-100k list from Alexa, we get a \emph{conservative} estimate of anycast adoption. To be even more conservative, we filter out anycast IP/24 addresses that are located in only two locations. We prefer to avoid false positives that might arise from wrong geolocation of the PlanetLab nodes. However, our subset is representative since it is very likely that the most popular websites in the Alexa ranking are also the ones attracting a significant number of users and adopt CDNs to handle the traffic volume.
\noindent
\fakesubsection{Simplicity.} Similarly, as a first step, we opt for simplicity and choose to consider only the IP/32 addresses for the landing pages in the Alexa list (e.g., google.com/, facebook.com/, wikipedia.com/). Because websites are complex, the next logical step is to obtain a comprehensive list of the IP addresses of all the servers contacted when users connect to a website. However, this would increase the list of IP/32 addresses to check, inflating the running time of the census.


\begin{table*}[t!]
\scriptsize
\centering
\begin{tabular}{|r|c|r
@{$\,\,$}c
@{$\,\,$}c
@{$\,\,$}c
@{$\,\,$}c
@{$\,\,$}c
@{$\,\,$}c|r|r|r|r|r|
c@{$\,\,$}
c@{$\,\,$}c|
}
\multicolumn{2}{c}{} &
\multicolumn{7}{c}{Location(*)} &
\multicolumn{5}{c}{} &
\multicolumn{3}{c}{Changes}\\
\hline
/24 subnet 		& Owner            	& no.& EU  & NA & SA & AS & AF & OC &  IP/32  & Vol. [GB]	& Flows [k]	&  Users   	  &  FQDN       & RTT & TTL & TTFB			\\ \hline
\hline
93.184.220.0	& EdgeCast-1   		&  7 & \C  &    &    &    &    &    & 112      	&  491 	& 10,579 	&  11,537 	  &  3,857		& & &  \\ 
68.232.35.0 	& EdgeCast-2      	& 20 & \C  & \C &    & \C & \C & 	 &  81      	&  578 	& 10,130 	&  11,414      &   764     	& & &  \\ 
68.232.34.0 	& EdgeCast-3      	& 7  & \C  &    &    &    &    &    &  46     	& 1307	& 4,257 	&  11,374      &   611	   	& & &  \\ 
93.184.221.0 	& EdgeCast-4     	& 7  & \C  &    &    &    &    &    &  54      	&  744 	& 2,279 	&  10,917      &  1,882 	& & &  \\ 
204.79.197.0 	& Microsoft     	& 53 & \C  & \C & \C & \C & \C & \C &   9      	&   14	& 1,166 	&   10655      &   149      & \C & \C & \C 	\\ 
178.255.83.0 	& Comodo   			& 3  & \C  & \C &    &    &    &    &   3      	&    2 	& 579 	    &   9,657      &    62      & & &  \\ 
108.161.189.0 	& NetDNA  			& 7  & \C  & \C &    &    &    &    &  83      	&   28 	& 471 	    &   8,861      &  1,988	    & \C & \C & \C 	\\ 
69.16.175.0 	& Highwinds-1    	& 9  & \C  & \C &    &    &    &    &   3      	&  620	& 1,398 	&   7,639      &   676 	    & & &  				\\ 
198.232.124.0 	& Unitas Global  	& 8  & \C  & \C &    &    &    &    &  61  	    &   29 	& 742 	    &   6,434      &  1,041     & \C & \C & \C 		\\ 
198.41.209.0 	& CloudFlare-1   	&16 &      & \C &    & \C &    & \C &  61      	&   86	& 224 	    &   4,063      &   352    	& & &  				\\ 
88.208.9.0 		& Advanced Hosters 	& 2 &      & \C &    &    &    &    &   4      	&  0.2	& 667 	    &   3,246      &    57      & \C & & \C 			\\ 
205.185.208.0 	& Highwinds-2  		&10 & \C   & \C &    &    &    &    &  21      	&   27	& 127 	    &   2,339      &    91	    & & &  				\\ 
198.41.247.0  	& CloudFlare-2   	&17 & \C   & \C & \C & \C &    & \C &  56      	&   19	& 48 	    &   1,242      &   117 	   	& & &  				\\ \hline
\multicolumn{1}{|c|}{Others}&	-	&- &	   &    &    &    &    &    &  12,420 	&  810 	& 11,151 	&  11,417      & 47,574		&& & 				\\ \hline
\multicolumn{1}{r}{\em Total} &	
\multicolumn{1}{c}{-}	&
\multicolumn{1}{r}{-} &
\multicolumn{6}{c}{} & 
\multicolumn{1}{r}{13,014}    & 
\multicolumn{1}{r}{4,762} &
\multicolumn{1}{r}{43,826} 	&  
\multicolumn{1}{r}{11,777}      & 
\multicolumn{1}{r}{58,397}	 \\

\end{tabular}
\\
\rule{0em}{1.1em}
(*) EU=Europe, NA=North America, SA=South America, AS=Asia, AF=Africa, OC=Oceania
\caption{Summary of results considering one week of traffic. \label{tab:summary}}
\end{table*}

\section{Passive  Characterisation}\label{sec:passive}

Having got a list of 328 IP/24 anycast subnets, we now leverage passive measurements to characterise the traffic they generate. The process is sketched in the bottom part of Fig.~\ref{fig:workflow}.
\subsection{Monitoring setup}
We instrumented a passive probe at one PoP of the operational network of an European country-wide ISP. The probe runs Tstat~\cite{finamore2011_tstat}, a passive monitoring tool that observes packets flowing on the links connecting the PoP to the ISP backbone network. Tstat rebuilds each TCP flow in real time, tracks it, and, when the connection is torn down, logs more than 100 statistics in a simple text file. 
For instance, Tstat logs the client and server IP addresses\footnote{We take care of obfuscating any privacy sensitive information in the logs. For instance customer IP addresses are anonymised using irreversible hashing functions, and only aggregate information are considered.
The deployment and the information collected for this has been approved by the ISP security and ethic boards.}, the application (L7) protocol type, the amount of bytes and packets sent and received, etc.
Tstat implements DN-Hunter~\cite{BermudezIMC2012}, a plugin that annotates each TCP flow with the server Fully Qualified Domain Name (FQDN) the client resolved via previous DNS queries. For instance, assume a client would like to access to \emph{www.acme.com}. It first resolves the hostname into IP/32 address(es) via DNS, getting 123.1.2.3. DN-Hunter caches this information. Then, when later at some time the same client opens a TCP connections to 123.1.2.3, DN-Hunter returns \emph{www.acme.com} from its cache and associate it to the flow. 
This is particularly useful for unveiling \emph{services} accessed from simple TCP logs.

For this study we leverage a dataset collected during the whole month of September 2014. 
It consists 2.0 billions of TCP flows being monitored, for a total of 270~TB of network traffic. 1.5 billion connections are due to web (HTTP or HTTPS) generating 199~TB of data.
We observe more than 20,000 customers IP addresses active over the month.\footnote{The anonymised customer IP address is an identifier of the household, which may hide several devices connected to the Internet through the same home gateway.}

Among the many measurements provided by Tstat, we consider for each TCP flow: 
(i) The Minimum Round-Trip-Time (RTT) between the Tstat probe and the server; 
(ii) the Minimum Time-To-Live (TTL) of packets sent by the server; 
(iii) the Time-To-First-Byte (TTFB), i.e., amount of time between the TCP SYN message and the first segment carrying data from the server;
(iv) the amount of downloaded bytes carried;
(v) the application layer protocol (e.g., HTTP, HTTPS, etc.); and
(vi) the FQDN of the server the client is contacting.
These metrics are straightforward to monitor, and details can be found in~\cite{finamore2011_tstat,BermudezIMC2012}. 

\subsection{A-CDN characterisation}

\begin{figure}[t!]
    \begin{center}
        \includegraphics[width=0.45\textwidth]{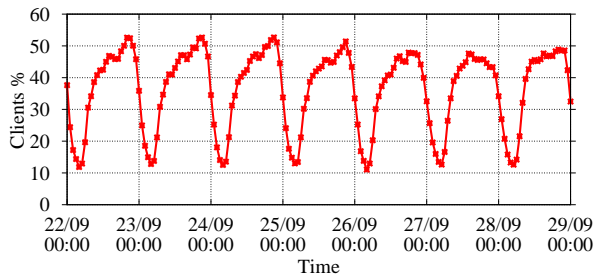}
		\caption{Percentage of customers that contact at least one A-CDN server in each 1h time bin.}
		\label{fig:any_client} 
    \end{center}
\end{figure}

By restricting our analysis on the IP/24 anycast subnets resulting from our census, we observe TCP traffic being served by some IP/32 addresses in those.\footnote{The logs report also UDP traffic to anycast addresses which Tstat identifies as DNS. We do not investigate more on this.}
Overall, almost 44 million TCP connections are managed by anycast servers. Those correspond to approximately 2\% of all web connections and 3\% of the aggregate HTTP and HTTPS volume,
for a total of 4.7~TB of data in the entire month. Definitively a not-negligible amount of traffic. All traffic is directed to TCP port 80 or 443, and labelled as HTTP and SSL/TLS by Tstat DPI. Only few exceptions are present, they represent in total about 0.26\% of all anycast traffic. These exceptions are mainly related to some ad-hoc protocols for multimedia streaming, email protocols, or DNS over TCP.
This testifies that today anycast is not anymore used for UDP services only, and A-CDNs are a reality. To corroborate this, Fig.~\ref{fig:any_client} shows evolution during one-week of the percentage of active customers that have encountered an A-CDN server during their normal web browsing activities. Besides exhibiting the classical day/night pattern, the figure shows that at peak time the probability to contact at least one A-CDN server is higher than 0.5.

Table~\ref{tab:summary} presents a summary of the results. It considers the IP /24 anycast networks that are contacted by more than 1000 customers. 
Due to space constrains we details only the top 13 most popular subnets with respect to the number of clients contacting them.
The remaining are aggregated as ``Others'' in the table.
For each subnet we list Owner, i.e., the organisations managing it as returned by Whois, and the number of locations found through our active probing methodology, along their presence in different continents. Interestingly, we observe well-known players like EdgeCast and CloudFlare, but also almost unknown companies offering A-CDN services to customers. Microsoft has its own A-CDN, which has the largest number of locations (53), offering a worldwide service. 
Notice how the number of locations is smaller than the one of traditional CDNs. For instance, EdgeCast largest IP/24 anycast subnet appears to have 20 different locations. Akamai CDN is instead known to have several thousands~\cite{Nygren:akamai}. While our census may have not located all possible caches locations, the two orders of magnitude of difference shows that A-CDNs are in their early deployment, and we expect the number of locations to grow in the future. For comparison, the Google Public DNS servers network for the 8.8.8.8 network has 55 worldwide resolvers~\cite{googleresolvers}.

\subsubsection{DNS Load Balancing}

An other interesting point to study is whether on owner offer DNS load balancing service or not. To study this aspect, we can not rely on passive measurements only since we have to understand if an FQDN has two or more distinct IP/32 addresses at the same time. By using \emph{Host} we discovered the number of distinct IP/32 addresses offered by each owner to each FQDN.

\subsection{Top A-CDN details}

\begin{figure*}[t!]
 \includegraphics[width=0.45\textwidth]{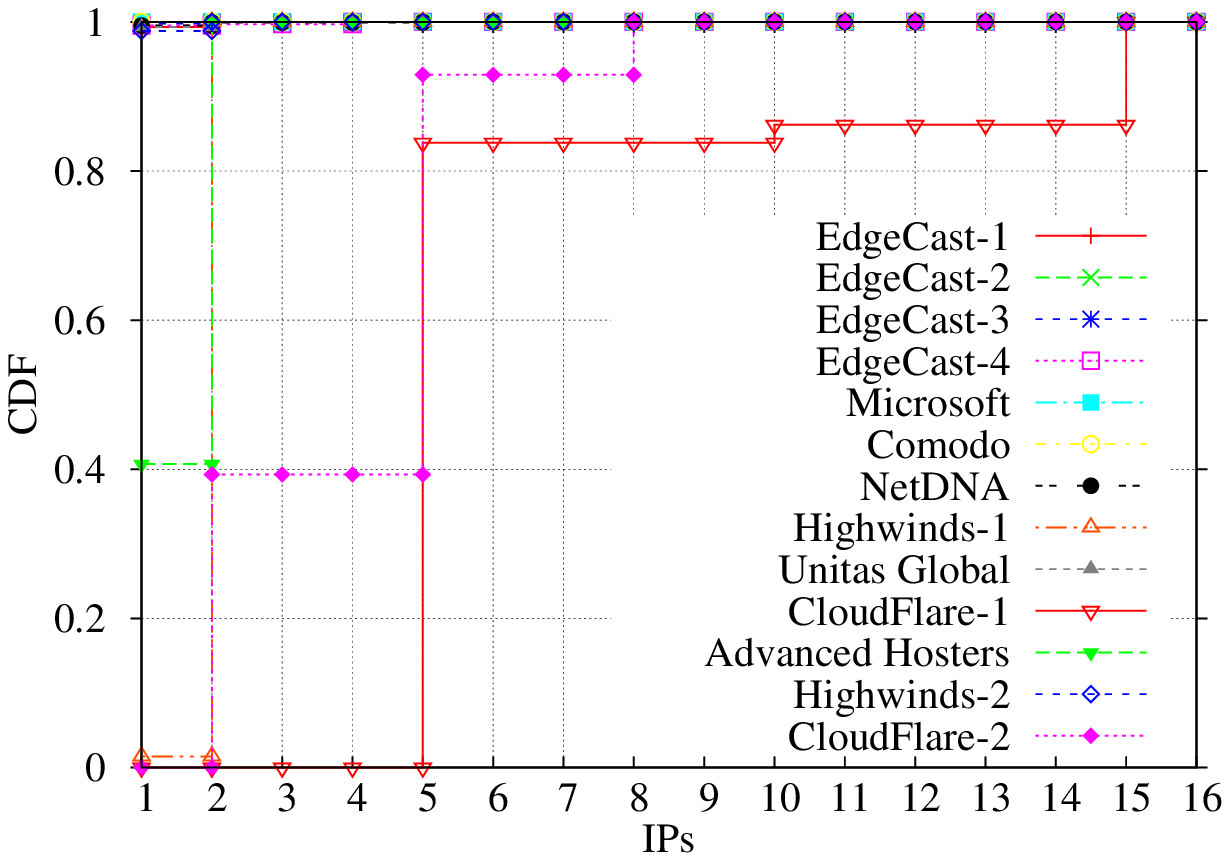}
 \includegraphics[width=0.45\textwidth]{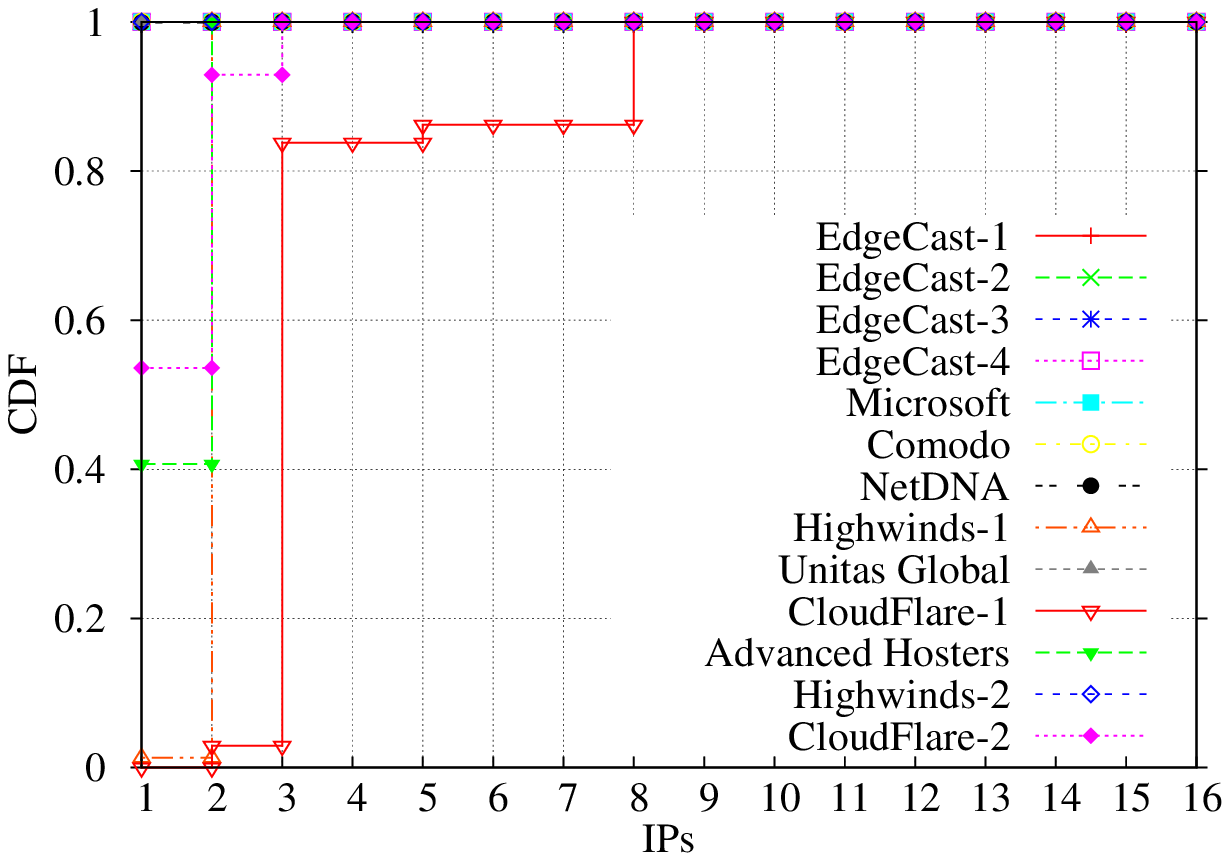}
\caption{All network vs Specific Network \label{fig:dns_load_balancing}
}
\end{figure*}

Aggregating one-week long Tstat logs, we detail information about volume and service offered.
For each IP/24 subnet, 
Table~\ref{tab:summary} reports the distinct A-CDN IP/32 addresses that have been contacted at least once, the total volume of bytes served, the number of flows, of users, and of distinct FQDNs. Interestingly, we observe a very heterogeneous scenario: the top four IP/24 networks are owned by EdgeCast, each serving more than 500~GB/month of traffic to more than 10,000 households.
Thousands of services (FQDNs) are involved. Those include very popular services, like Wordpress, Twitter, Gravatar, Tumblr, Tripadvisor, Spotify, etc. Each FQDN is uniquely resolved to the same IP/32 address (but the same IP/32 address serves multiple FQDNs).
Interestingly, this behaviour is shared among most of the studied A-CDNs, meaning that they do not rely on DNS for load-balancing. An exception is given by CloudFlare's networks which offer DNS load-balancing. Indeed we saw that CloudFlare offer up to 8 IP/32 addresses in the same IP/24 for the same FQDN. By look at the left graph of Figure \ref{fig:dns_load_balancing} we can see the Cumulative Distribution Function (CDF) of the number of distinct IPs/32 addresses eployed for each FQDN divided by owner. In this first graph we evaluated the number of IP/32 addressees without take into account to which subnet/24 they belong. E.g., some FQDNs of CloudFlare-1 have also IP/32 addressees belonging to CloudFlare-2. As we can see CloudFlare (1 or 2) use more than 1 IP/32 addresses for all their FQDNs. The same situation is present by considering only the IP/32 belonging to the same subnet of the owner, as depicated in the left part of the figure. As we can see here the maximum number of distinct IP/32 addresess drop from 15 to 8 for CloudFlare-1 while IP/32 hadled with a single IP/32 address became 54.4\% for CloudFlare-2.

Microsoft directly manages its own A-CDN. We discovered 53 locations, where only 9 IP addresses are present. Those serve Bing, Live, Msn, and other Microsoft.com services. Since it handles quite a small amount of data and flows, we checked if there are other IP/32 servers handling those popular Microsoft service in the logs. We found indeed that all of \emph{bing.com} pages and web searches are served by the Microsoft A-CDN, while static content such as pictures, map tiles, etc. are actually retrieved by Akamai CDN. This suggests that Microsoft is using both a traditional CDN and its own A-CDN at the same time.

Next is Comodo. It focuses its business on serving certificate validations via OCSP services: lot of customers uses it to fetch little information. Only 3 IP/32 addresses have been active from our passive vantage point. Note that servers have been located only in Europe and North America. 

Highwinds A-CDNs instead supports video services for advertisement companies, and images for popular adult content websites. Notice the relative longer lived content (more data, fewer flows). CloudFlare A-CDN serves both popular website like Reddit, and less known services, like specialised forums. 
A detailed list of the top-10 services for each of the 13 top networks in table \ref{tab:summary} is available in appendix \ref{sec:services}.

\begin{figure}[t!]
    \begin{center}
        \includegraphics[width=0.45\textwidth]{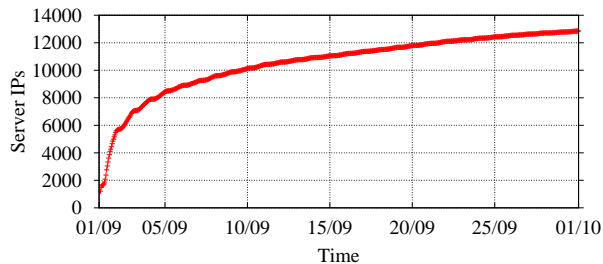}
		\caption{Number of distinct IP/32 encountered during time.}
		\label{fig:server_growth} 
    \end{center}
\end{figure}

Other A-CDN providers are present, which serve several tens of thousands of services. In total, 13,014 IP addresses have been found active during the whole month. Fig.~\ref{fig:server_growth} details the discovery process by reporting the number of unique IP/32 addresses discovered over time in the entire month of September 2014. As shown, the discovery quickly grows during the first days, then the process slows down. Surprisingly, after 30 days, the growth is still far from being complete.

\section{Routing anomalies}
\label{sec:affinity}

\begin{figure*}[t!]
 \centering
 \hspace{-1em}
 \includegraphics[width=0.33\textwidth]{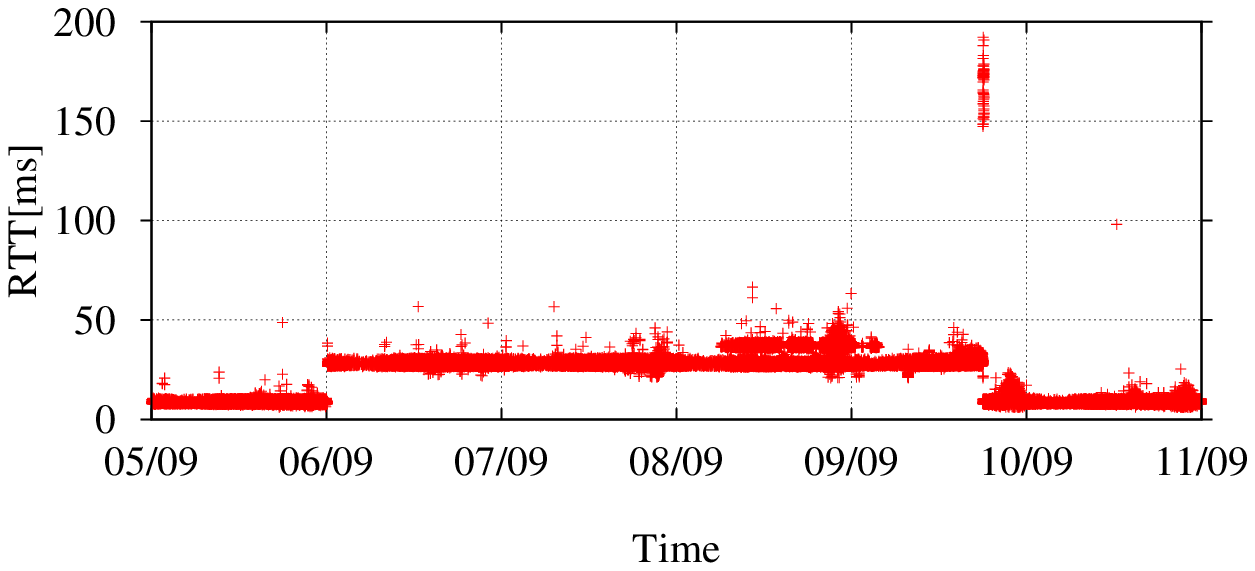}
 \includegraphics[width=0.33\textwidth]{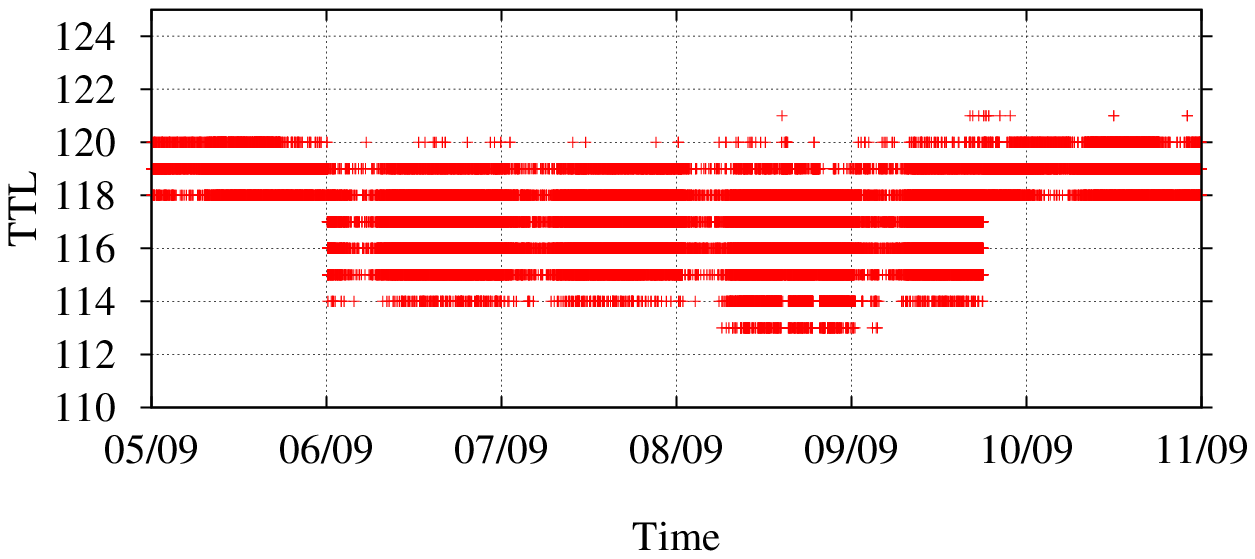}
 \includegraphics[width=0.33\textwidth]{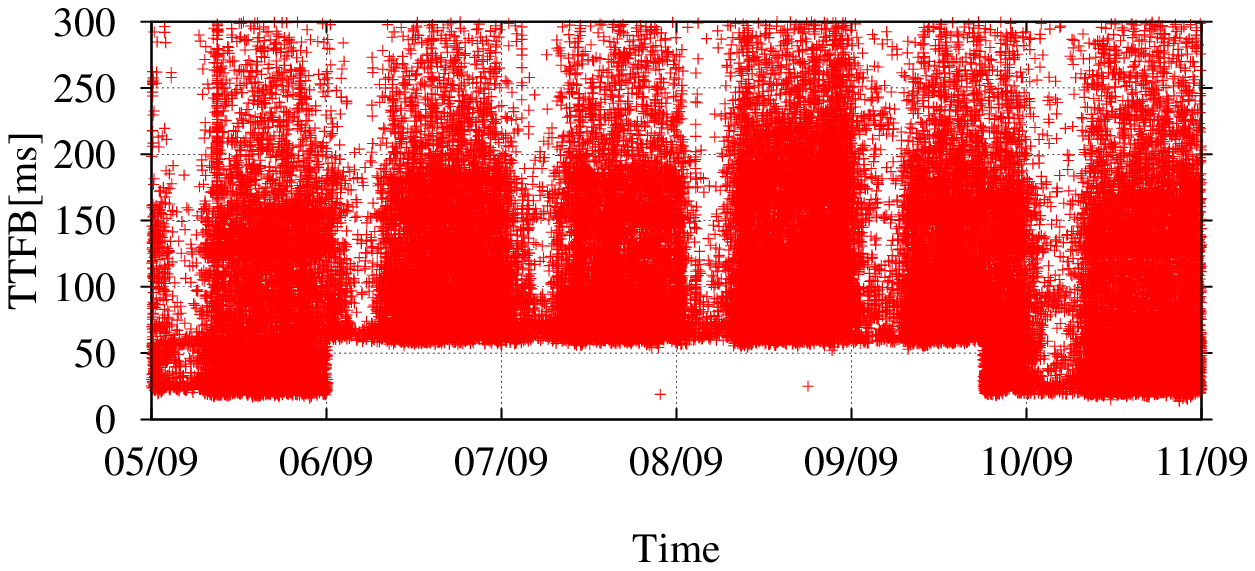}
\\
\hspace{-1em}
 \includegraphics[width=0.33\textwidth]{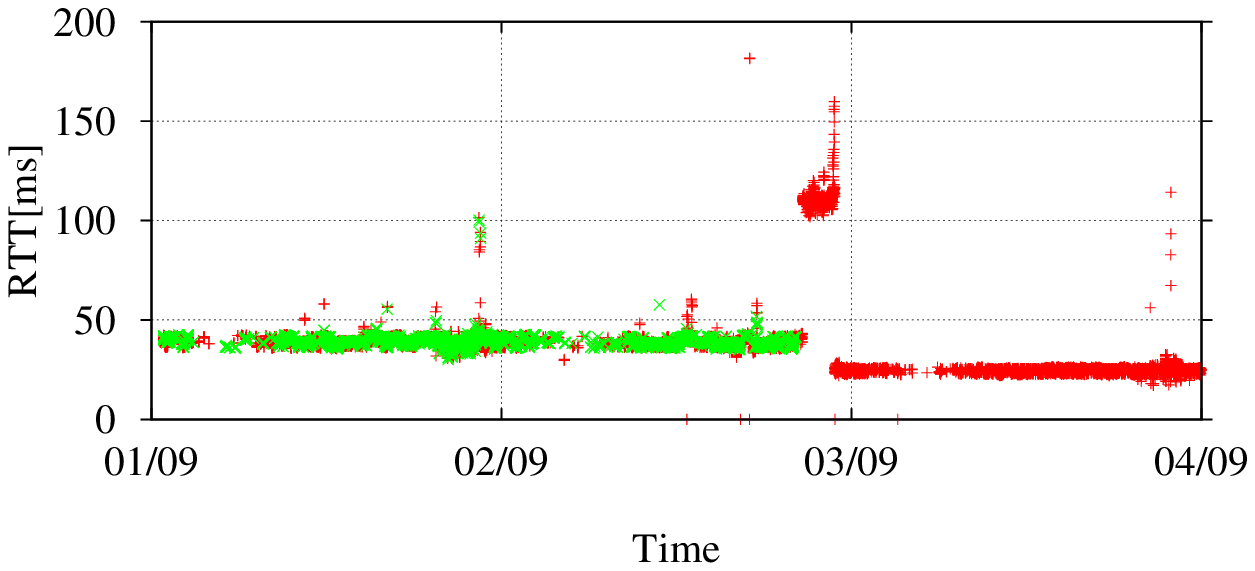}
 \includegraphics[width=0.33\textwidth]{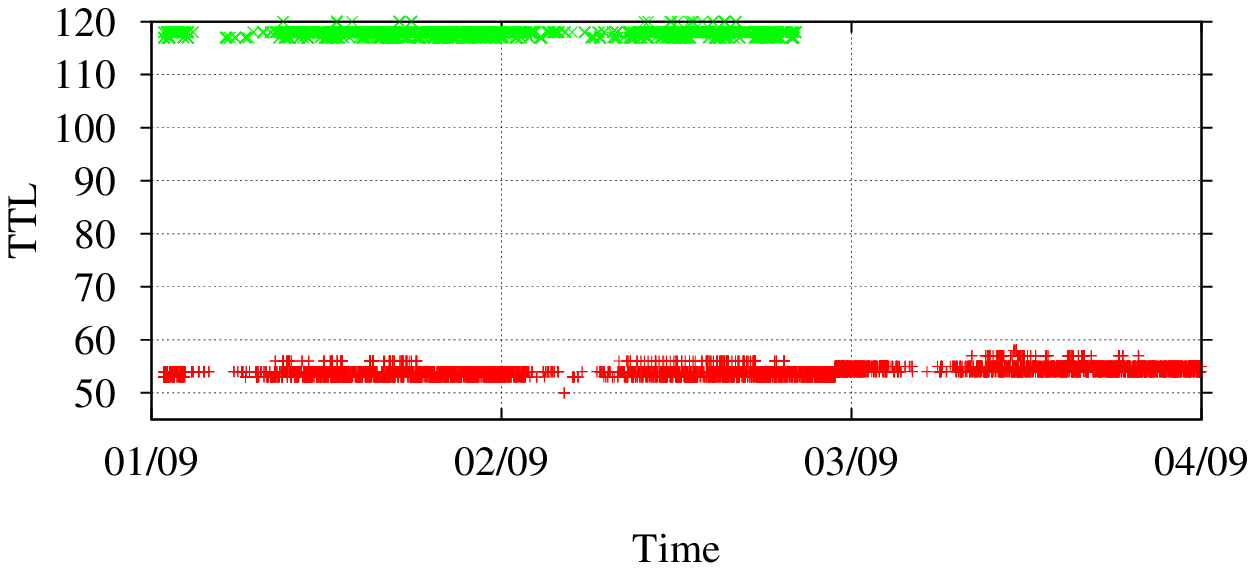}
 \includegraphics[width=0.33\textwidth]{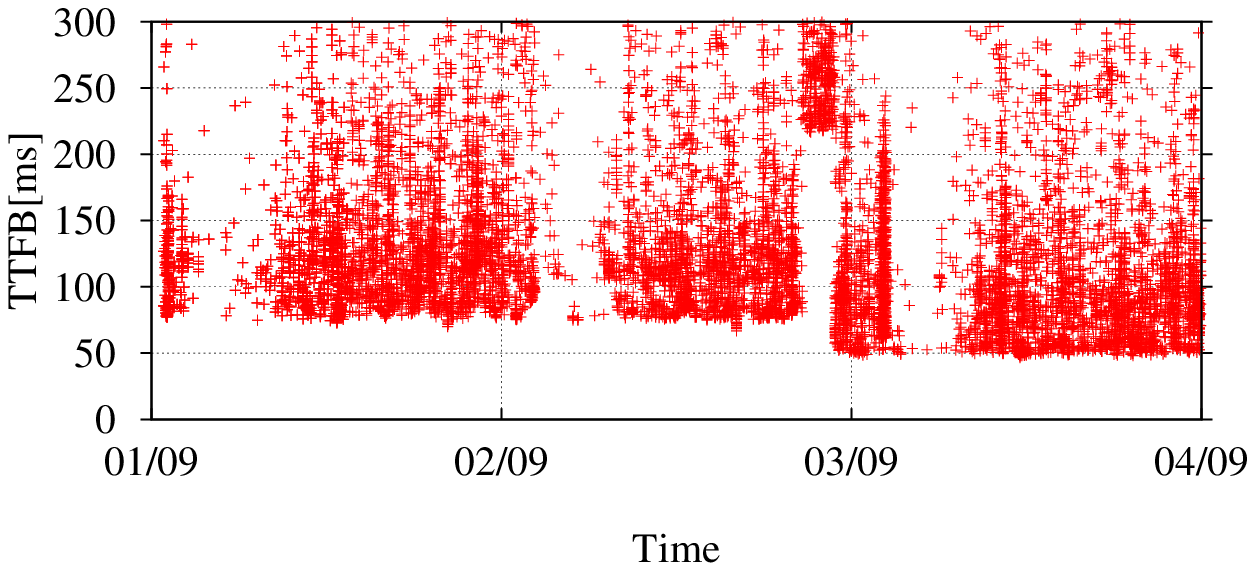}
 \\
\caption{Microsoft (top), NetDNA (bottom) event details}
\label{fig:events}
\end{figure*}

One of the popular belief about Internet routing is that the paths may change quite frequently due to faults, misconfigurations, peering changes, and, for A-CDNs, voluntary load-balancing optimisation. Thus, anycast services are mostly suitable for datagram services based on UDP, while stateful and connection-oriented services using TCP may suffer troubles due to sudden nearest server changes that may cause abrupt interruption of ongoing TCP connections, and loss of consistency on state. In this section we thus look for evidences that hint for possible routing changes.
In particular, we look at changes in IP TTL, TCP RTT and Time To First Byte that may suggest of possible path changes for a given IP/24 subnet\footnote{Routing changes may cause changes in the data we observe. However, we have no means to confirm them and we leave to the reader to drive a conclusion.}.
We consider the whole month of September 2014 dataset.
The last column of Table~\ref{tab:summary} reports our findings: almost surprisingly, we observe that for the majority of cases we observe no notable change during the entire month. Not reported here due to lack of space, this is testified by a practically constant RTT, identical pattern for TTL and TTFB through the entire month.

There are four events that we believe it is worth reporting to illustrate some of the changes that we observe. Fig.~\ref{fig:events} reports the detail of the evolution of the RTT, TTL and TTFB for two events. Each dot is a  measure for a single TCP flow.
Top plot refers to Microsoft A-CDN from the 5$^{th}$ to the 11$^{th}$ of September 2014.
Focus on the RTT first. It exhibits a sudden change at midnight of the 5$^{th}$, when the RTT jumps from 8~ms to 28~ms. It then goes back three days later, after a transient phase during which the RTT gets higher than 150~ms. Similar changes are observed in the TTL, where two patterns are clearly visible. We argue the multiple values of the TTL are due to different servers being contacted \emph{inside} the internal IP/24 network of the datacenter (and thus reached through different number of internal routers). The TTFB clearly shows the impact on performance when a further location is contacted. The variability in the TTFB depends on multiple factors, including browser pre-opening TCP connections, and server processing time. However, the minimum TTFB is clearly constrained by (twice) the RTT. 

Bottom plots in Fig.~\ref{fig:events} show a similar event for NetDNA A-CDN: during September 2$^{nd}$, the RTT first jumps to 110~ms, then to 23~ms. This corresponds to a change in the TTL and in the TTFB patterns as well.
Interestingly, the TTL patterns suggests the presence of servers that use different initial values of the TTL: a group of servers chooses 128 (in green), while another group uses 64 (in red). When on the late evening of the 3$^{rd}$ of September the routing changes, traffic is routed to a likely different cache, where all servers pick 64 as initial TTL value, i.e., we observe the green dots to suddenly disappear.
Changes for the other two events are similar and not reported here for the sake of brevity.


In summary, while we observe changes in the anycast path to reach the A-CDN caches, those events are few, and each different routing configuration last for days. This is different from the patterns shows by traditional CDNs, where load balancing changes are more frequent~\cite{Finamore:youtube,finamore_publicdns}. This can be related to the also moderately smaller number of locations, and to the different load balancing policies A-CDN providers are adopting.
Clearly, a longer study is needed to better quantify the routing changes over time.

\section{Conclusions and Discussion}\label{sec:open}

We presented in this paper a first characterisation of Anycast-enabled CDN. Starting from a census of anycast subnets, we analysed passive measurements collected from an actual network to observe the usage and the stability of the service offered by A-CDNs. Our finding unveil that A-CDNs are a reality, with several players adopting anycast for load balancing, and with users that access service they offer on a daily basis. Interestingly, passive measurements reveal a very stable service, with stable paths and cache affinity properties.
In summary, anycast is increasingly used, A-CDNs are prosperous and technically viable. 

This work is far from yielding a complete picture, and it rather raises a number of interesting questions that we list in the following to stimulate discussion in the community:
\noindent
\fakesubsection{Completeness}. We have so far focused on a subset of the anycast IPv4 space. It follows that results conservatively estimate anycast usage, but this also means that more effort is needed to build (and especially maintain) an Internet-wide anycast census. Similarly, dataset spanning over larger period of time and related to more vantage points can enable more general results with respect to the actual characteristics of A-CDNs.
%
%
\noindent
\fakesubsection{Horizontal comparison with IP unicast}. Albeit very challenging, efforts should be dedicated to compare Unicast vs Anycast CDNs for modern web services. To the very least, a statistical characterisation of the pervasiveness of the deployments (e.g., in term of RTT) and its impact on objective measures (e.g., time to the first byte, average throughput, etc.) could be attempted. However, many more vantage points than the single one considered in this work would be needed to gather statistically relevant samples from the user population viewpoint.
\noindent
\fakesubsection{Vertical investigation of CDN strategies}. From our initial investigation, we noticed radically different strategies, with e.g., hybrid DNS resolution of few anycast IP addresses, use of many DNS names mapping to few anycast IPs, use of few names mapping to more than one anycast IPs, etc. Gathering a more thorough understanding of load balancing in these new settings is a stimulant intellectual exercise which is not uncommon in our community.
\noindent
\fakesubsection{Further active/passive measurement integration}. As anycast replicas are subject to BGP convergence, a long-standing myth is that it would forbid use of anycast for connection-oriented services relying on TCP. Given our results, this myth seems no longer holding. Yet, while we did not notice in our time frame significant changes in terms of IP-level path length, more valuable information would be needed from heterogeneous sources, and by combining active and passive measurements.

%

%

\section*{Acknowledgements}

This work has been partly carried out at LINCS.
The research leading to these results has received funding from the European Union under the FP7 Grant Agreements no.\,318627 (Integrated Project ``mPlane'').

\begin{small}
\bibliographystyle{plain}

\end{small}

\appendix
\section{Service Table}  \label{sec:services}

In this section we report the top 10 services of each of the top 13 networks reported in table \ref{tab:summary}.  Each service is defined by its \textit{second level domain} e.g., either www.bing.com or bing.it will be bing only. For each services than we report the following data related to the traffic generated during the month of September 2014: the number of distinct servers that served during the month it, the volume in MB, the number of flows, the number of distinct users who requested the service and finally the distinct number of FQDNs e.g.,  www.bing.com or bing.it count as two. It is important to remark that the number of distinct servers can be different with respect to the DNS load balancing policy explained in section X. Here number of IPs can be greater since it is evaluated in the whole month instead of a precise moment. Therefore, a service might be moved from one server to an other not for load-balancing reason but due to maintenance.

\begin{table}[h!]
\scriptsize
\centering
\begin{tabular}{|l|c|c|c|c|c|}
\hline
\multicolumn{6}{|c|}{93.184.220.0/24 Edgecast-1}\\  \hline
Service                                  &       Servers  &  Vol.[MB]  &  Flows    &  Users  &  FQDN  \\  \hline
digicert                                 &       1        &  1662.75   &  1079308  &  10661  &  5     \\  \hline
wp                                       &       1        &  36905.8   &  834071   &  9076   &  9     \\  \hline
bkrtx                                    &       1        &  1011.81   &  153314   &  8883   &  2     \\  \hline
optimizely                               &       1        &  6077.63   &  264662   &  8688   &  1     \\  \hline
crwdcntrl                                &       1        &  1038.8    &  150670   &  8682   &  1     \\  \hline
omniroot                                 &       1        &  16420     &  66105    &  8420   &  2     \\  \hline
w55c                                     &       2        &  361.442   &  100788   &  7804   &  2     \\  \hline
typekit                                  &       1        &  5559.58   &  165720   &  7675   &  3     \\  \hline
edgecastcdn                              &       55       &  21959.4   &  281715   &  5923   &  291   \\  \hline
sascdn                                   &       1        &  1337.82   &  56819    &  5794   &  2     \\  \hline
\end{tabular}
\end{table}

\begin{table}[h!]
\scriptsize
\centering
\begin{tabular}{|l|c|c|c|c|c|}
\hline
\multicolumn{6}{|c|}{68.232.35.0/24 Edgecast-2}\\  \hline
Service                                 &       Servers  &  Vol.[MB]  &  Flows    &  Users  &  FQDN  \\  \hline
twitter                                 &       1        &  45665.3   &  4042986  &  11254  &  4     \\  \hline
gravatar                                &       1        &  5218.38   &  1387880  &  9961   &  5     \\  \hline
twimg                                   &       4        &  113178    &  1142388  &  9442   &  14    \\  \hline
adrcdn                                  &       1        &  1234.27   &  95104    &  7796   &  1     \\  \hline
tiqcdn                                  &       1        &  782.397   &  120563   &  6897   &  1     \\  \hline
edgecastcdn                             &       33       &  5867.74   &  162551   &  4871   &  63    \\  \hline
doublepimp                              &       1        &  1231.38   &  106590   &  4732   &  6     \\  \hline
tumblr                                  &       5        &  305513    &  898342   &  4612   &  34    \\  \hline
exoclick                                &       1        &  670.502   &  214197   &  4442   &  2     \\  \hline
bstatic                                 &       2        &  20377.3   &  520660   &  4177   &  2     \\  \hline
\end{tabular}
\end{table}

\begin{table}[h!]
\scriptsize
\centering
\begin{tabular}{|l|c|c|c|c|c|}
\hline
\multicolumn{6}{|c|}{68.232.34.0/24 Edgecast-3}\\  \hline
Service                                 &       Servers  &  Vol.[MB]  &  Flows    &  Users  &  FQDN  \\  \hline
microsoft                               &       1        &  24739.8   &  1731861  &  9708   &  11    \\  \hline
msecnd                                  &       3        &  120398    &  776628   &  8890   &  216   \\  \hline
adrcdn                                  &       1        &  81589     &  179684   &  8286   &  1     \\  \hline
aspnetcdn                               &       1        &  5865      &  117402   &  7649   &  1     \\  \hline
tripadvisor                             &       1        &  12020.5   &  221128   &  4738   &  1     \\  \hline
w55c                                    &       1        &  761.561   &  33623    &  4557   &  1     \\  \hline
msn                                     &       1        &  1467.56   &  134112   &  4509   &  10    \\  \hline
mozilla                                 &       1        &  31728.3   &  105097   &  3513   &  12    \\  \hline
phncdn                                  &       1        &  456726    &  113367   &  2630   &  2     \\  \hline
edgecastdns                             &       2        &  239.702   &  7705     &  1753   &  2     \\  \hline
\end{tabular}
\end{table}

\begin{table}[h!]
\scriptsize
\centering
\begin{tabular}{|l|c|c|c|c|c|}
\hline
\multicolumn{6}{|c|}{93.184.221.0/24 Edgecast-4}\\  \hline
Service                                  &       Servers  &  Vol.[MB]  &  Flows   &  Users  &  FQDN  \\  \hline
weborama                                 &       1        &  5265.13   &  287956  &  9232   &  4     \\  \hline
jwpcdn                                   &       2        &  5100.11   &  409004  &  8979   &  3     \\  \hline
longtailvideo                            &       1        &  1126.79   &  109453  &  6221   &  11    \\  \hline
ad4mat                                   &       1        &  322.927   &  67397   &  4824   &  5     \\  \hline
webads                                   &       1        &  422.79    &  35335   &  4115   &  2     \\  \hline
mozilla                                  &       1        &  31575.6   &  96193   &  3376   &  1     \\  \hline
deviantart                               &       2        &  17367.9   &  55189   &  2782   &  11    \\  \hline
edgecastcdn                              &       20       &  47051.5   &  49753   &  2392   &  170   \\  \hline
ppstatic                                 &       1        &  2790.91   &  98391   &  2229   &  5     \\  \hline
everyplay                                &       1        &  11286.9   &  18813   &  1861   &  1     \\  \hline
\end{tabular}
\end{table}

\begin{table}[h!]
\scriptsize
\centering
\begin{tabular}{|l|c|c|c|c|c|}
\hline
\multicolumn{6}{|c|}{204.79.197.0/24 Microsoft}\\  \hline
Service                                  &       Servers  &  Vol.[MB]   &  Flows   &  Users  &  FQDN  \\  \hline
bing                                     &       2        &  10067.4    &  840891  &  10501  &  41    \\  \hline
microsoft                                &       1        &  1315.81    &  173235  &  4540   &  1     \\  \hline
live                                     &       6        &  1758.57    &  44898   &  907    &  37    \\  \hline
windowssearch                            &       2        &  221.675    &  13149   &  893    &  2     \\  \hline
a-msedge                                 &       6        &  157.061    &  8899    &  264    &  34    \\  \hline
msn                                      &       3        &  306.819    &  8572    &  250    &  9     \\  \hline
akadns                                   &       4        &  18.0363    &  611     &  72     &  17    \\  \hline
myhomemsn                                &       1        &  40.4178    &  1919    &  56     &  1     \\  \hline
msnrewards                               &       1        &  3.77615    &  207     &  38     &  1     \\  \hline
livefilestore                            &       1        &  0.0163116  &  2       &  1      &  1     \\  \hline
\end{tabular}
\end{table}
\begin{table}[h!]
\scriptsize
\centering
\begin{tabular}{|l|c|c|c|c|c|}
\hline
\multicolumn{6}{|c|}{178.255.83.0/24 Comodo}\\  \hline
Service                                  &       Servers  &  Vol.[MB]  &  Flows   &  Users  &  FQDN  \\  \hline
comodoca                                 &       2        &  1678.81   &  389377  &  9390   &  3     \\  \hline
usertrust                                &       2        &  246.683   &  113390  &  7859   &  3     \\  \hline
netsolssl                                &       2        &  24.9833   &  10629   &  2456   &  2     \\  \hline
gandi                                    &       2        &  26.6111   &  12705   &  2395   &  3     \\  \hline
trust-provider                           &       2        &  16.4111   &  17395   &  1948   &  2     \\  \hline
terena                                   &       2        &  30.2878   &  7075    &  939    &  3     \\  \hline
csctrustedsecure                         &       2        &  4.34491   &  5984    &  677    &  3     \\  \hline
globessl                                 &       2        &  1.79976   &  3017    &  197    &  3     \\  \hline
incommon                                 &       2        &  0.917895  &  336     &  134    &  3     \\  \hline
ssl                                      &       2        &  0.509459  &  266     &  132    &  4     \\  \hline
\end{tabular}
\end{table}

\begin{table}[h!]
\scriptsize
\centering
\begin{tabular}{|l|c|c|c|c|c|}
\hline
\multicolumn{6}{|c|}{108.161.189.0/24 NetDNA}\\  \hline
Service                                   &       Servers  &  Vol.[MB]  &  Flows  &  Users  &  FQDN  \\  \hline
addtoany                                  &       1        &  541.206   &  57956  &  5404   &  1     \\  \hline
jquerytools                               &       1        &  315.579   &  16239  &  3761   &  1     \\  \hline
netdna-cdn                                &       43       &  4121.45   &  80414  &  3385   &  530   \\  \hline
buysellads                                &       1        &  283.627   &  21806  &  2088   &  2     \\  \hline
popcash                                   &       1        &  8.78808   &  10245  &  1775   &  2     \\  \hline
netdna-ssl                                &       16       &  1418.63   &  30571  &  1599   &  84    \\  \hline
flowplayer                                &       2        &  52.0184   &  5622   &  1238   &  2     \\  \hline
fastcdn                                   &       1        &  20.8619   &  5571   &  681    &  1     \\  \hline
feedbackify                               &       1        &  10.9618   &  3347   &  568    &  1     \\  \hline
chitika                                   &       2        &  5.09706   &  1758   &  546    &  2     \\  \hline
\end{tabular}
\end{table}
\begin{table}[h!]
\scriptsize
\centering
\begin{tabular}{|l|c|c|c|c|c|}
\hline
\multicolumn{6}{|c|}{69.16.175.0/24 Highwinds-1}\\  \hline
Service                                 &       Servers  &  Vol.[MB]  &  Flows   &  Users  &  FQDN  \\  \hline
hwcdn                                   &       2        &  106703    &  84692   &  3282   &  72    \\  \hline
adxpansion                              &       2        &  9567.55   &  165124  &  3041   &  2     \\  \hline
xvideos                                 &       2        &  248916    &  155720  &  2603   &  29    \\  \hline
sexad                                   &       2        &  3217.49   &  36290   &  1918   &  1     \\  \hline
reporo                                  &       2        &  2285.14   &  19235   &  1679   &  2     \\  \hline
adjuggler                               &       2        &  120.766   &  36262   &  1396   &  4     \\  \hline
camads                                  &       2        &  2733.45   &  23726   &  1183   &  1     \\  \hline
sancdn                                  &       2        &  2146.91   &  10736   &  1156   &  1     \\  \hline
crossrider                              &       2        &  337.945   &  22467   &  1107   &  4     \\  \hline
nsimg                                   &       2        &  31150.4   &  17954   &  1089   &  2     \\  \hline
\end{tabular}
\end{table}

\begin{table}[h!]
\scriptsize
\centering
\begin{tabular}{|l|c|c|c|c|c|}
\hline
\multicolumn{6}{|c|}{198.232.124.0/24 Unitas Global}\\  \hline
Service                                   &       Servers  &  Vol.[MB]  &  Flows   &  Users  &  FQDN  \\  \hline
netdna-cdn                                &       28       &  10276.5   &  54751   &  1600   &  355   \\  \hline
datafastguru                              &       5        &  282.8     &  283056  &  1486   &  4     \\  \hline
cedexis                                   &       2        &  145.48    &  6378    &  1462   &  2     \\  \hline
mdotm                                     &       2        &  526.938   &  3829    &  1128   &  2     \\  \hline
pusher                                    &       1        &  31.99     &  3486    &  674    &  1     \\  \hline
petametrics                               &       1        &  130.947   &  8736    &  597    &  1     \\  \hline
ad-score                                  &       1        &  250.462   &  70998   &  597    &  1     \\  \hline
engageya                                  &       2        &  707.5     &  94965   &  519    &  4     \\  \hline
revcontent                                &       1        &  351.279   &  4839    &  503    &  1     \\  \hline
rnbjunk                                   &       1        &  270.919   &  14527   &  480    &  5     \\  \hline
\end{tabular}
\end{table}
\begin{table}[h!]
\scriptsize
\centering
\begin{tabular}{|l|c|c|c|c|c|}
\hline
\multicolumn{6}{|c|}{198.41.209.0/24 CloudFlare-1}\\  \hline
Service                                  &       Servers  &  Vol.[MB]  &  Flows  &  Users  &  FQDN  \\  \hline
reddit                                   &       8        &  587.413   &  48806  &  3308   &  46    \\  \hline
redditstatic                             &       5        &  158.645   &  17850  &  1237   &  1     \\  \hline
redditmedia                              &       10       &  571.371   &  36644  &  629    &  7     \\  \hline
cursecdn                                 &       3        &  5934.29   &  42142  &  449    &  41    \\  \hline
camplace                                 &       3        &  328.093   &  9234   &  444    &  12    \\  \hline
pluginnetwork                            &       3        &  756.187   &  12587  &  306    &  6     \\  \hline
gfycat                                   &       3        &  6556.39   &  2949   &  275    &  9     \\  \hline
comodo                                   &       3        &  25323.2   &  11266  &  212    &  1     \\  \hline
smugmug                                  &       3        &  487.699   &  1776   &  187    &  107   \\  \hline
diablofans                               &       3        &  247.697   &  2507   &  135    &  4     \\  \hline
\end{tabular}
\end{table}

\begin{table}[h!]
\scriptsize
\centering
\begin{tabular}{|l|c|c|c|c|c|}
\hline
\multicolumn{6}{|c|}{88.208.9.0/24 Advanced Hosters}\\  \hline
Service                                &       Servers  &  Vol.[MB]  &  Flows   &  Users  &  FQDN  \\  \hline
xhcdn                                  &       2        &  159.63    &  531141  &  2911   &  10    \\  \hline
vstreamcdn                             &       2        &  9.57007   &  29015   &  1128   &  2     \\  \hline
ahcdn                                  &       4        &  4.07057   &  13341   &  770    &  2     \\  \hline
mystreamservice                        &       2        &  5.2034    &  17473   &  467    &  3     \\  \hline
wildcdn                                &       1        &  4.47277   &  13075   &  450    &  1     \\  \hline
alotporn                               &       2        &  1.4562    &  5054    &  430    &  1     \\  \hline
vipstreamservice                       &       1        &  0.898319  &  2393    &  173    &  1     \\  \hline
tryboobs                               &       2        &  0.394236  &  1516    &  165    &  1     \\  \hline
inxy                                   &       2        &  0.417435  &  1205    &  105    &  1     \\  \hline
ohsesso                                &       1        &  1.03889   &  2425    &  100    &  1     \\  \hline
\end{tabular}
\end{table}
\begin{table}[h!]
\scriptsize
\centering
\begin{tabular}{|l|c|c|c|c|c|}
\hline
\multicolumn{6}{|c|}{205.185.208.0/24 Highwinds-2}\\  \hline
Service                                   &       Servers  &  Vol.[MB]  &  Flows  &  Users  &  FQDN  \\  \hline
hwcdn                                     &       7        &  11952.7   &  58551  &  1126   &  50    \\  \hline
thestaticvube                             &       1        &  11281     &  15686  &  745    &  4     \\  \hline
vidible                                   &       1        &  700.347   &  2883   &  548    &  1     \\  \hline
trustedshops                              &       1        &  11.8071   &  1761   &  197    &  1     \\  \hline
xplosion                                  &       1        &  13.9726   &  4113   &  90     &  3     \\  \hline
chzbgr                                    &       1        &  192.167   &  707    &  53     &  3     \\  \hline
bose                                      &       1        &  127.601   &  475    &  43     &  2     \\  \hline
brainient                                 &       1        &  186.67    &  298    &  32     &  4     \\  \hline
metartnetwork                             &       1        &  2.53812   &  1470   &  29     &  1     \\  \hline
blaze                                     &       1        &  18.7321   &  262    &  25     &  1     \\  \hline
\end{tabular}
\end{table}

\begin{table}[h!]
\scriptsize
\centering
\begin{tabular}{|l|c|c|c|c|c|}
\hline
\multicolumn{6}{|c|}{198.41.247.0/24 CloudFlare-2}\\  \hline
Service                                  &       Servers  &  Vol.[MB]  &  Flows  &  Users  &  FQDN  \\  \hline
filmstream                               &       1        &  1452.82   &  9080   &  490    &  2     \\  \hline
sendapplicationget                       &       1        &  27.8062   &  6305   &  275    &  7     \\  \hline
mangaeden                                &       1        &  13713.2   &  11517  &  157    &  5     \\  \hline
feedly                                   &       3        &  221.195   &  14278  &  114    &  4     \\  \hline
mobisystems                              &       1        &  17.3486   &  215    &  55     &  3     \\  \hline
keep2share                               &       1        &  5.18796   &  368    &  36     &  4     \\  \hline
mafa                                     &       1        &  121.536   &  436    &  34     &  3     \\  \hline
racing-games                             &       1        &  32.6952   &  192    &  23     &  2     \\  \hline
switchfly                                &       2        &  31.5114   &  229    &  22     &  1     \\  \hline
ofreegames                               &       1        &  48.0902   &  171    &  21     &  2     \\  \hline
\end{tabular}
\end{table}

\end{document}